\documentclass[dvips]{article}
\usepackage{icrctc07}

\title{On the relation between the proton-air cross section and fluctuations of the shower longitudinal profile}
\shorttitle{Proton-air cross section from logitudinal profiles}
\authors{R. Ulrich$^1$, J. Bl\"umer$^{1,2}$, R. Engel$^1$, 
F. Sch\"ussler$^1$ and M. Unger$^1$}
\shortauthors{R. Ulrich et al.}
\afiliations{
$^1$ Institut f\"ur Kernphysik, Forschungszentrum Karlsruhe, 
Postfach 3640, 76021 Karlsruhe, Germany\\
$^2$ Universit\"{a}t Karlsruhe, Institut f\"{u}r Experimentelle Kernphysik, 76128 Karlsruhe, Germany
}
\email{Ralf.Ulrich@ik.fzk.de}

\abstract{The current status and prospects of deducing the proton-air
  cross section from fluorescence telescope measurements of extensive
  air showers are discussed.  As it is not possible to observe the
  point of first interaction, $X_ {\rm 1}$, directly, other
  observables closely linked to $X_{\rm 1}$ must be inferred from the
  measured longitudinal profiles. This introduces a dependence on the
  models used to describe the shower development.  Systematic
  uncertainties arising from this model dependence, from the
  reconstruction method itself and from a possible non-proton contamination
  of the selected shower sample are discussed.}

\begin{document}
\maketitle

\section{Introduction}

Indirect cosmic ray measurements by means of extensive air shower
(EAS) observations are difficult to interpret. Models needed for a
deeper understanding of the data have to be extrapolated over many
decades in energy. This is the case for high energy (HE) interaction
models, but also applies to the primary composition of cosmic
rays. Unfortunately a changing primary composition and changes in the
HE interaction characteristics can have similar effects on EAS
development and are difficult to separate. \\
One of the key parameters for EAS development is the cross section
$\sigma_{\rm p-air}$ of a primary proton in the atmosphere.
Of course, only the part of the cross section leading to secondary particle production
is relevant for EAS development, which we call for simplicity here $\sigma_{\rm p-air}$.
But also the
production cross section contains contributions which cannot be observed in EAS.
As diffractive interactions of primary particles with air
nuclei do not (target dissociation) or weakly (projectile dissociation)
influence the resulting EAS, any measurement based on EAS is
insensitive to these interactions. Therefore, we define an 
effective cross section to require an inelasticity $k_{\rm inel}=1-\frac{E_{\rm
    max}}{E_{\rm tot}}$ of at least 0.05
\begin{equation}
  \sigma_{\rm p-air}^* = \sigma_{\rm p-air}(k_{\rm inel}\ge0.05).
\end{equation}
In the following the amount of traversed matter before an interaction with
$k_{\rm inel}\ge0.05$ is called $X_{\rm 1}$.
Taking this into account the reconstructed value of $\sigma_{\rm
  p-air}^*$ needs to be altered by a model dependent correction 
$\sigma^{\rm model}_{\rm p-air}(k_{\rm inel}<0.05)$. This correction amounts
to 2.4~\% for SIBYLL \cite{sibyll}, 3.9~\% for QGSJETII.3 \cite{qgsjetII}
and 5.5~\% for QGSJET01 \cite{qgsjet01}, resulting in a model uncertainty of
$\sim$3~\%. 

All EAS simulations are performed in the CONEX \cite{conex}
framework. To account for the limited reconstruction accuracy of a realistic EAS detector, 
$X_{\rm max}$ is folded with an Gaussian function having 
20~gcm$^{-2}$ width, which corresponds roughly to the resolution of the Pierre Auger
Observatory \cite{AugerXmaxResolution}.

\section{\boldmath $X_{\rm max}$-distribution ansatz}

The most prominent source of shower fluctuations is the interaction path length of the
primary particle in the atmosphere. However the EAS development itself
adds a comparable amount of fluctuations to observables like $X_{\rm
  max}$. This is mainly due to the shower startup phase, where the EAS
cascade is dominated by just a few particles. 
Our approach to fit the full distribution of $X_{\rm max}$ does therefore
handle the primary interaction point explicitly and the EAS development in a parametric
way
\begin{eqnarray}
  \frac{dP}{dX_{\rm max}^{\rm exp}}  & = & \int dX_{\rm max}
  \int dX_{\rm 1} \; \frac{e^{-X_{\rm 1}/{\lambda^{*}_{\rm p-air}}}}{{\lambda^{*}_{\rm
  p-air}}}  \nonumber \\
  &\times& P_{\Delta X}(\Delta X+{X_{\rm shift}}, \lambda^{*}_{p-air})\nonumber \\
  &\times& P_{X_{\rm max}}(X_{\rm max}^{\rm exp}-X_{\rm max}),
\label{eq:folding}
\end{eqnarray}
where $\Delta X$ was introduced as $X_{\rm max}-X_{\rm 1}$.  Thus the
$X_{\rm max}$-distribution is written as a double convolution, with
the first convolution taking care of the EAS development and the
second convolution handling the detector resolution. In this model we
have two free parameters $\lambda^{*}_{\rm p-air}$, which is directly
related to $\sigma^{*}_{\rm p-air}$, and $X_{\rm shift}$, needed to
reduce the model dependence. Note that Eq.~(\ref{eq:folding}) differs
from the HiRes approach \cite{Belov} and that used in the simulation studies in \cite{Weihei}
by explicitly including the cross section dependence in $P_{\Delta X}$.

\begin{figure}[t]
  \includegraphics[width=\linewidth]{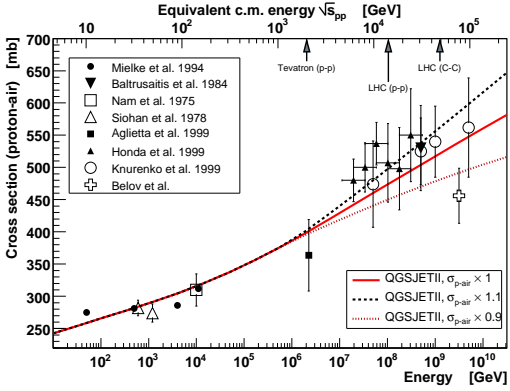}
  \caption{Impact of a 10~\% change of $\sigma_{\rm p-air}$ in QGSJETII 
    at 10~EeV. Data from \cite{Belov,Mielke,Aglietta,Baltrusaitis,Knurenko,Honda,Gaisser}.}
  \label{f:conex}
\end{figure}
\begin{figure}
  \includegraphics[width=\linewidth]{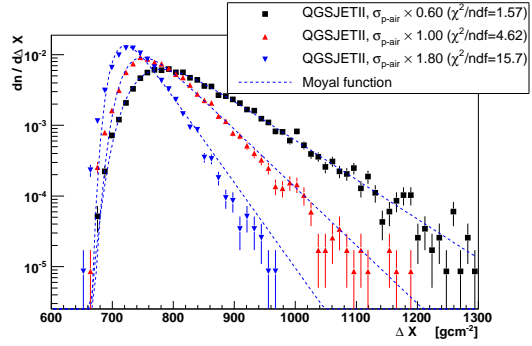}
  \caption{Example fits of Eq.~(\ref{e:moyal}) to simulated $P_{\Delta
      X}$-distributions at 10~EeV. }
  \label{f:CXfits}
\end{figure}

\begin{figure}[b!]
  \includegraphics[width=\linewidth]{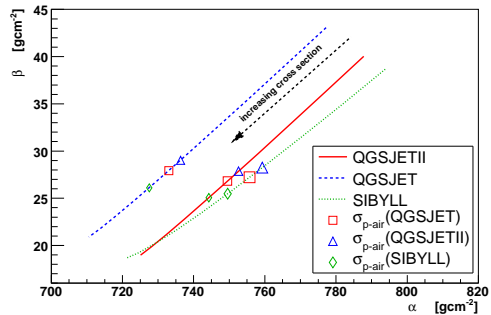}
  \caption{Resulting $\sigma_{\rm p-air}$-dependence of the parametrized 
    $P_{\Delta X}$-distribution. The markers denote the location of the
    original HE model cross sections.} 
  \label{f:CX2D}
\end{figure}

The simulated $P_{\Delta X}$-distributions can be parametrized efficiently with
the Moyal function
\begin{equation}
  P_{\Delta X}(\Delta X) = \frac{e^{-\frac{1}{2}(t +
  e^{-t})}}{\beta\sqrt{2\pi}} \;\;\textrm{and}\;\; t=\frac{\Delta X -
  \alpha}{\beta} 
  \label{e:moyal}
\end{equation}
using the two free parameters $\alpha$ and $\beta$.

\section{Impact of \boldmath $\sigma_{\rm p-air}$ on EAS development}

To include the cross section dependence of $P_{\Delta X}$ in a cross
section analysis at 10~EeV, we modified CONEX for several HE models
such that the cross section used in the simulation is replaced by
\begin{equation}
  \sigma_{\rm p-air}^{\rm modified}(E)=\sigma_{\rm p-air}(E)\cdot \left(1+f(E)\right),
\end{equation}
with the energy dependent factor $f(E)$, which is equal to $0$ for $E\le$1~PeV
and
\begin{equation}
  f(E) = (f_{\rm 10EeV}-1)\cdot\frac{\log_{10}(E/1\;\textrm{PeV})}{\log_{10}(1\;\textrm{EeV}/1\;\textrm{PeV})}
\end{equation}
for $E>1$~PeV, reaching $f_{\rm 10EeV}$ at $E=10$~EeV. This modification
accounts for the increasing uncertainty of $\sigma_{\rm p-air}$ for
large energies (see Fig.~\ref{f:conex}). Below 1~PeV (Tevatron energy),
$\sigma_{\rm p-air}$ is predicted within a given HE model by fits to the
measured $p\bar p$ cross section.\\ 
The cross section dependence of $P_{\Delta X}$ and the corresponding
parametrizations are shown in Fig.~\ref{f:CXfits}. At large $\Delta
X$, the simulated distributions are not perfectly reproduced by the
parametrizations. This effect worsens for large cross sections, as can be
observed from the increasing $\chi^2/ndf$ (see Fig.~\ref{f:CXfits}). Also the
deviation of the Moyal function from the $P_{\Delta X}$-distribution depends
on the HE model. It is biggest for QGSJETII and smallest for
SIBYLL. Unfortunately this disagreement produces a systematic overestimation
of $\sim$30~mb for the reconstructed $\sigma_{\rm p-air}$. This is visible in
all the following results and will be addressed in future work by making the
parametrization more flexible.\\
The dependence of $\alpha$ and $\beta$ on $\sigma_{\rm p-air}$ can be interpolated with
a polynomial of 2nd degree. Fig.~\ref{f:CX2D} gives an overview of
this interpolation in the $\alpha$-$\beta$ plane. Obviously the
$P_{\Delta X}$ predicted by different HE model are not only a
consequence of the different model cross sections.

\section{Results}

\begin{figure}
  \includegraphics[width=\linewidth]{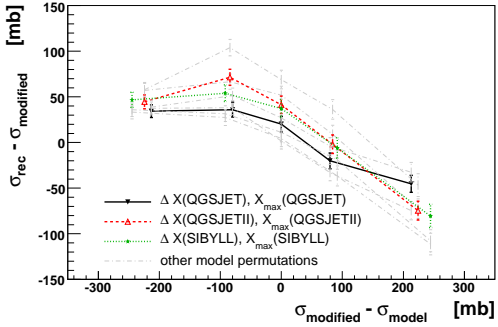}
  \vspace*{-.6cm}
  \caption{Sensitivity and HE model dependence of the $\sigma_{\rm p-air}$ reconstruction for a pure
    proton composition at 10~EeV.}
  \label{f:proton}
\end{figure}

\noindent
{\bf Pure proton composition}\\ 
In Fig.~\ref{f:proton} we show the reconstructed $\sigma^{\rm rec}_{\rm p-air}$ for
simulated showers with modified high energy model cross section ,
$\sigma^{\rm modified}_{\rm p-air}$. The original HE cross section
$\sigma^{\rm modified}_{\rm p-air}-\sigma^{\rm model}_{\rm p-air}=0$ can be 
reconstructed with a statistical uncertainty of $\sim$10~mb,
whereas the uncertainty caused by the HE models is about $\pm$50~mb.
At smaller cross sections the reconstruction results in  a slight
overestimation ($<50$~mb). But for larger cross sections there occurs
a significant underestimation of the input cross section. 
This is mainly due to the worse description of $P_{\Delta X}$ by the used
Moyal function for large values of $\sigma_{\rm p-air}$ (see last section). 

\begin{figure}
  \includegraphics[width=\linewidth]{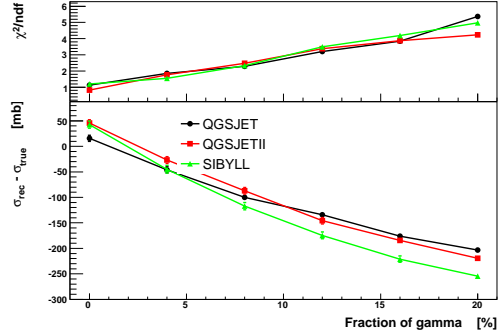}
  \vspace*{-.6cm}
  \caption{Systematic caused by photon primaries at 10~EeV.}
  \label{f:photon}
\end{figure}

\noindent
{\bf Photon primaries}\\ 
Primary photons generate deeply penetrating showers. Even a small
fraction of photon showers has a noticeable effect on the tail of the
$X_{\rm max}$-distribution \cite{Weihei}. Fig.~\ref{f:photon}
demonstrates how much a few percent of photons could influence the
reconstructed $\sigma_{\rm p-air}$. The current limit on the photon
flux is 2~\% at 10~EeV \cite{AugerPhotonLimit}. Note that there is a
clear trend of an increasing $\chi^2/ndf$ with increasing photon fraction,
meaning the photon signal is not compatible with the proton model.

\begin{figure}[t]
  \includegraphics[width=\linewidth]{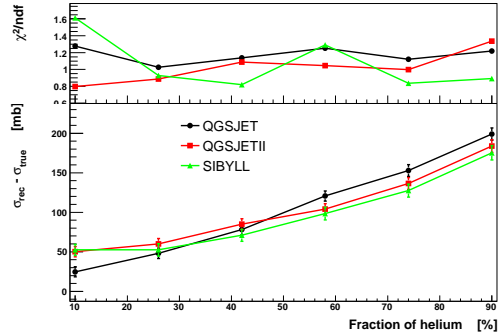}
  \vspace*{-.6cm}
  \caption{Systematic caused by helium primaries at 10~EeV.}
  \label{f:helium}
\end{figure}

\noindent
{\bf Helium primaries}\\ 
On the contrary, helium induced EAS are very similar to proton
showers. Therefore their impact on $\sigma_{\rm p-air}$ is significant
and very difficult to suppress, see
Fig.~\ref{f:helium}. Interestingly, even for large helium contributions
there is no degradation of the quality of the pure proton model fit ($\chi^2/ndf$ is
flat).  Thus it is not possible in a simple way to distinguish between
a 25\%~proton / 75\%~helium mixture or just a pure proton composition
with a cross section increased by about 150~mb.

\begin{figure}
  \includegraphics[width=\linewidth]{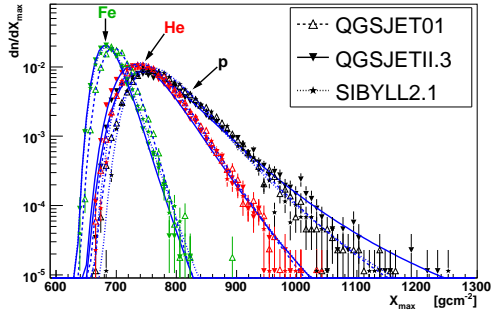}
  \caption{Composition impact on $X_{\rm max}$ at 10~EeV.}
  \label{f:comp}
\end{figure}

\noindent
{\bf Outlook: Mixed primary composition}\\
Fluctuations and the mean value of the $X_{\rm max}$-distribution are
frequently utilized to infer the composition of primary cosmic rays
\cite{AugerXmax}. It is well 
understood how nuclei of different mass $A$ produce shower maxima 
at different depth $X_{\rm max}(A)$ and how shower-to-shower fluctuations
decrease with $A$ (semi-superposition model). \\
The relative change of the $X_{\rm max}$-distribution from a pure proton to a
pure mass $A$ primary composition can be evaluated using CONEX. To fit $X_{\rm
  max}$-distributions we use the formula \cite{john}
\begin{eqnarray}
  \label{e:john}
  \frac{dP}{dX_{\rm max}}(A) & = &  N \cdot e^{-\left(\frac{\sqrt{2}(X_{\rm
  max}-X_{\rm peak})}{\gamma\cdot(X_{\rm max}-X_{\rm peak}+3\cdot\delta)}\right)^{2}}
\nonumber\\
\end{eqnarray}
with four parameters $N$, $X_{\rm peak}$, $\gamma$ and $\delta$. The
normalization constant $N$ was not fitted, but set to reproduce the known
number of events. Fig.~\ref{f:comp} shows how the $X_{\rm
  max}$-distributions for proton, helium and iron primaries are positioned
relative to each other for several HE models. This relative alignment can be
utilized during $\sigma_{\rm p-air}$-fits to reduce the composition
dependence.%
The total mixed composition $X_{\rm max}$-distribution is
then the weighted sum of the individual primaries
\begin{equation}
  \frac{dP}{dX_{\rm max}^{\rm mix}}(X_{\rm max}) = \sum_i
  \omega_i \; \frac{dP}{dX_{\rm max}}(A_i, X_{\rm max})
\end{equation}
where the weights $\omega_i$ are additional free parameters to be fitted
together with $X_{\rm shift}$ and $\lambda^{*}_{\rm p-air}$. The shape of
$\frac{dP}{dX_{\rm max}}(A)$ for $A > 1$ is always assumed to change relative
to the proton distribution.

First studies indicate that the correlation between the reconstructed
composition and the corresponding $\sigma_{\rm p-air}$ does not allow a
measurement of the cross section. The situation is expected to be more
promising if the parameter $X_{\rm shift}$ is fixed, however, the model
dependence of the analysis will then be larger than shown here.

\bibliography{icrc1027}
\bibliographystyle{unsrt}

\end{document}